# Learning to Build Shapes by Extrusion

T. V. Christiansen[1], K. Pandey[2], A. Reinders[1], K. Singh[2], M. R. Hannemose[1], J. A. Bærentzen[1]

[1]Technical University of Denmark, Kgs. Lyngby, Denmark
[2]University of Toronto, Toronto, Canada

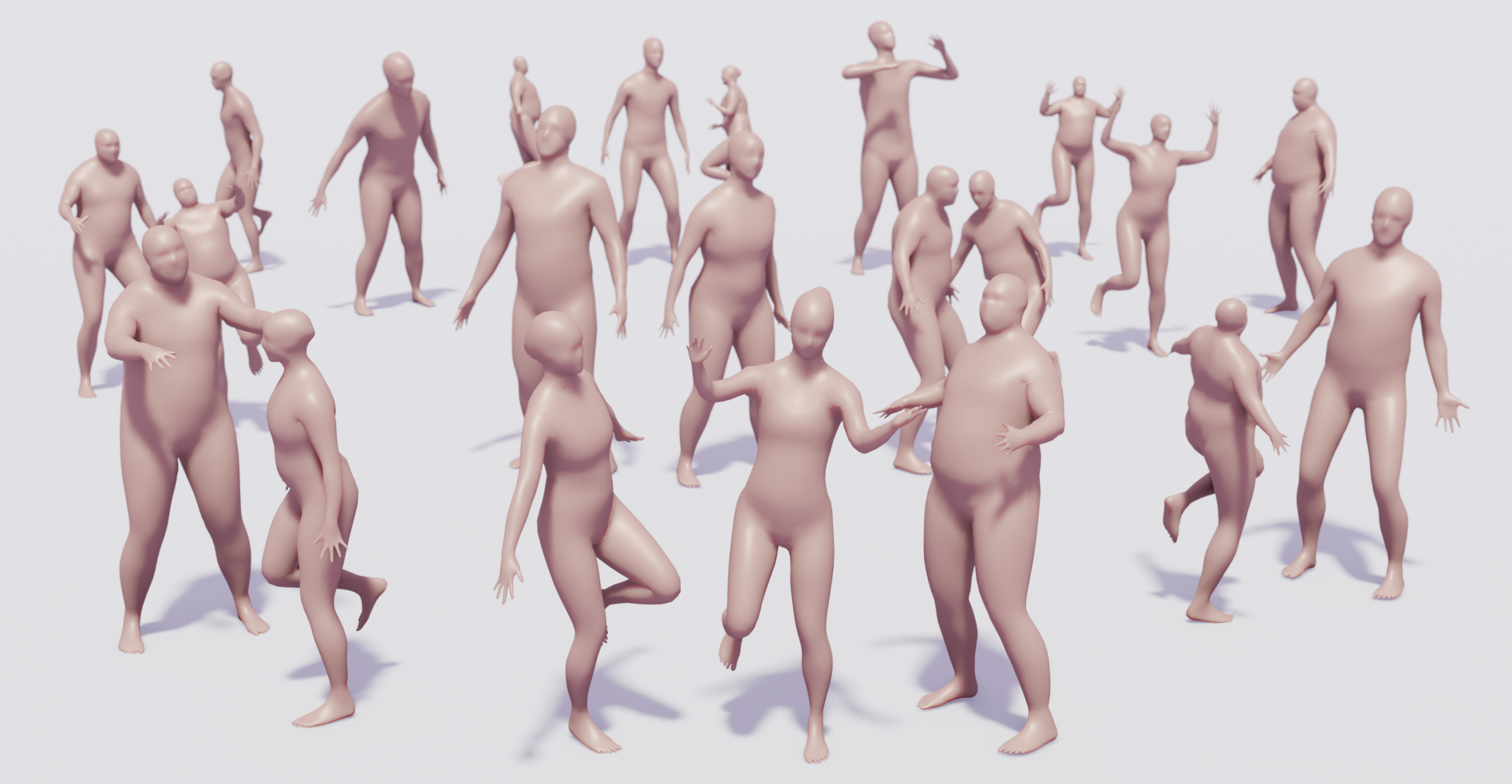

**Figure 1:** *A mashup of humans created from LLM generated TEE sequences and then built using our quad dominant framework*

**Abstract**

*We introduce Text Encoded Extrusions (TEE), a text-based representation that expresses mesh construction as sequences of face extrusions rather than polygon lists, and a method for generating 3D meshes from TEE using a large language model (LLM). By learning extrusion sequences that assemble a mesh, similar to the way artists create meshes, our approach naturally supports arbitrary output face counts and produces manifold meshes by design, in contrast to recent mesh generative transformer based models. The learnt extrusion sequences can also be applied to existing meshes - enabling editing in addition to generation. To train our model, we decompose a library of quadrilateral meshes with non-self-intersecting face loops into constituent loops, which can be viewed as their building blocks, and finetune an LLM on the steps for reassembling the quadrilateral meshes by performing a sequence of extrusions. We demonstrate that our representation enables reconstruction, novel shape synthesis, and the addition of new features to existing meshes.*

## 1. Introduction

Learning-based models for 3D shape generation have achieved incredible results over the last few years [PFS*19,PJBM23,LGT*23, LTZ*24], generating a wide variety of detailed 3D shapes. However, most of these methods rely on implicit surface representations, which struggle with learning sharp features, as these are inherently discontinuous and can only be approximated. Moreover, the methods often produce overly dense triangle meshes when extracting surfaces from neural fields, as a byproduct of iso-contouring methods such as the Marching Cubes [LC87] algorithm and its variants [Gib98, CTFZ22, SMH*23]. This is a problem, as a compact 3D mesh is often the desired output for downstream applications like computer games or video animations. To tackle these issues, a recent line of work based on transformers [VSP*17] has been proposed. Similar to natural language processing with transformers, these methods generate the primitives of a mesh sequentially, thereby learning the mesh directly rather than an implicit representation. This is advantageous, as artists often desire control over the structure of the polygonal mesh, something that is not easily provided by an iso-surface. On the other hand, these recent methods produce relatively coarse meshes [TLH*24, CWL*24]. By representing every single mesh primitive as a token (whether it be the vertex positions and/or the polygonal faces), the sequence length of a moderately high-resolution mesh becomes critically long since most transformer models have a limited context window, which, in turn, is because the attention mechanism often scales quadratically with the sequence length. In addition, transformers typically generate a discrete probability distribution over output tokens to predict the next token. This makes it difficult for them to predict continuous quantities, such as the coordinates of mesh vertices. As a result, many current methods [SAA*24, CWL*24, GSLD24] restrict the vertex positions to a predefined spatial grid. Finally, because shapes are generated sequentially, many methods only produce a soup of primitives rather than a manifold mesh.

To address these challenges, we propose learning to build a mesh through a sequence of extrusions rather than directly learning the sequence of polygons that form the mesh. We thereby cast the problem of explicit mesh generation as a matter of mesh representation instead of network architecture design. Our method is thus able to reproduce entire meshes, explore variations, and modify existing shapes. Rather than devising a network architecture for learning sequences of extrusions, we convert the extrusion commands to text, Text Encoded Extrusions (TEE), and learn sequences of extrusions through fine-tuning an existing off-the-shelf large language model (LLM). Our approach has several advantages: we impose no limit on the level of detail of the output mesh, the extrusion operations ensuring that we always obtain a mesh with the expected face-loop connectivity; face-loop extrusion operations produce meshes that inherently capture the flow-line structure of objects, which is crucial for deformation and other downstream applications [BWSS12]; we are able to use our extrusion framework to augment existing meshes; and we are free to choose any LLM.

In summary, our main contributions are:

- A learning-based methodology for generating 3D meshes via sequences of extrusions.
- Employing a Large Language Model for 3D mesh generation.
- Auto-completing features on meshes

We publish our code on GitHub and release two new datasets: quadrilateral versions of the triangle meshes from the DFAUST [BRPMB17] and MANO [RTB17] datasets.

## 2. Related work

Generative methods in computer graphics have been an active research field for many years. Before the rise in popularity of learning-based methods, geometric models were generated in a variety of ways, including L-systems [Pru86], evolutionary methods [Sim94], and model synthesis [MM11]. However, we focus on learning-based methods in the following, and we divide this class of methods into four different groups:

**Neural implicits:** More recently, neural implicit surface representations have attracted significant attention [PFS*19, TLY*21, MST*22, CMPM20, LWJ*22, GSF22, PJBM23]. In this type of representation, a neural network is trained to approximate a scalar and/or vector field, such as the signed distance field of a shape whose surface is implicitly defined as a level set of the learned function. From this, a mesh can be extracted using, e.g., Marching Cubes [LC87]. This type of representation is very versatile but struggles to capture thin, sharp features, as a bend with high curvature is a discontinuity, which can only be approximated. Moreover, a very high grid resolution is needed to extract a mesh with edges of high curvature, which typically leads to dense meshes even in regions with sparse features. Although this problem can be remedied using mesh simplification algorithms, it is still a non-trivial task to detect which parts of the mesh to simplify.

**Point cloud conditioned methods:** Another line of research focuses on generating 3D shapes indirectly. These methods create a polygonal surface by conditioning on a point cloud, which can be generated using a diffusion model [NJD*22, ZVW*22, SLL*24, HRLL24]. For instance, [NJD*22] [NJD*22] and [ZVW*22] [ZVW*22] generate a point cloud, convert it to an implicit surface, and hence a mesh. In SpaceMesh [SLL*24], a transformer is used to predict the polygonal faces of a mesh that matches a given point cloud. The method generates a manifold mesh by construction since it is based on the halfedge data structure. However, the method is limited by its memory consumption and cannot generate meshes of more than 2000 vertices. Similar to Spacemesh, EdgeRunner [TLH*24] is an autoregressive model able to generate a mesh by conditioning on a point cloud or an image. They make use of a more compact and efficient mesh representation, which allows them to reduce the number of tokens per face. This enables them to generate meshes with up to 4,000 triangle faces. An extension of this work is Meshtron [HRLL24], which also generates meshes based on an input point cloud. Meshtron adopts a new Hourglass Transformer architecture that allows for generated meshes with up to 64,000 triangle faces.

**Unconditional mesh generation:** In a number of recent works, meshes are generated unconditionally by sampling autoregressively from a transformer. These works include Polygen [NGEB20], MeshGPT [SAA*24], MeshXL [CCP*24] and MeshAnything V2 [CWL*24]. Some of these can operate conditionally, but we focus on unconditional generation since this is more pertinent. The methods differ in their network architectures and in how they encode meshes into token sets, but otherwise they build on the same principles. A mesh is predicted autoregressively, one primitive at a time, and space is discretized into a coarse grid that stores the vertices of the mesh. This is an artifact of the transformer network's training, which outputs a probability distribution over a discrete set of tokens, as true continuous predictions perform poorly. Moreover, by predicting only one primitive at a time, the methods do not guarantee that the output meshes are manifold - unlike our method. Finally, all methods can only handle meshes with fewer than $1,600$ triangles. If EdgeRunner and Meshtron are included, the capacity increases to $4,000$ triangle faces, as there is no published code for Meshtron [HRLL24] yet.

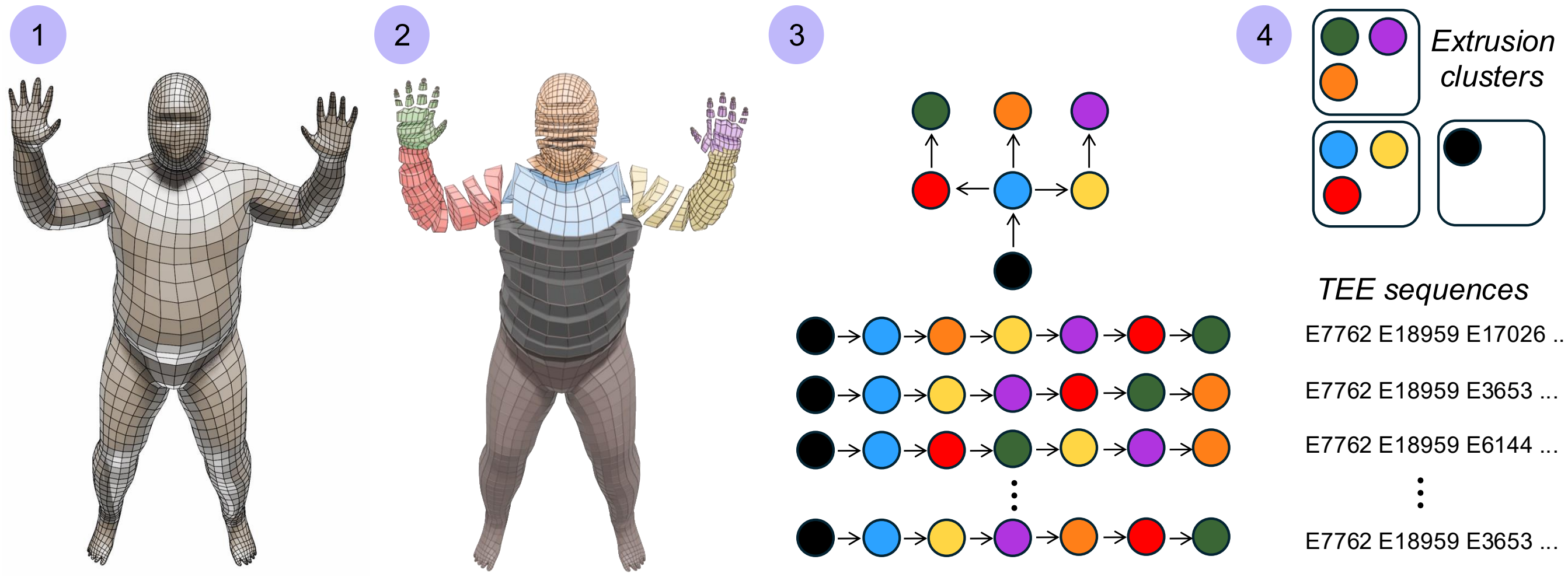

**Figure 2:** ***Method overview:*** *1) Our method takes as input an FEQ mesh. 2) A user selects a face-loop and all face-loops above that are found. 3) From the decomposition process of the FEQ mesh, a DAG of extrusion operations is obtained, and this is topologically sorted to provide different ways of re-assembling the mesh. 4) The extrusions operations are clustered according to geometric similarity, and together with the DAG-orderings, TEE sequences can be made. An LLM is then fine-tuned on the TEE sequences and from this able to generate new sequences, which can be converted to meshes.*

While these methods are the ones that most closely resemble our work, they differ significantly because we do not generate meshes at the primitive level but at the component level. This enables us to bypass the use of a 3D grid and makes the method insensitive to output resolution. In Table 1, we compare our method against neural implicit methods and methods that generate meshes explicitly using transformers.

**Table 1:** *Comparison of capabilities across implicit methods, previous transformer-based mesh generators (Explicit generation), and our approach.*

| | Implicit | Explicit generation | Ours |
|---|---|---|---|
| Sharp features | Non trivial | ✓ | ✓ |
| Max number of faces | No limit | $\leq 4,000$ [TLH*24] | No limit |
| Guaranteed manifold | ✓ | ✗ | ✓ |
| Continuous verts | ✓ | ✗ | ✓ |
| Feature completion | Non trivial | ✗ | easy |
| Good flow structure | Non trivial | ✗ | ✓ |

**Component level generation:** LegoGPT [PDL*25], which produces shapes in units of blocks, is tangentially related to our work. Their method takes a 3D mesh, voxelizes it on a coarse grid to create a LEGO design, and then trains an LLM to assemble the shape from LEGO bricks. Despite the similarity in building shapes from components, our work is fundamentally different since we operate on meshes with unconstrained vertex positions. Furthermore, our extrusions adapt to the target patch, unlike LEGO bricks. The transformer-based tree generation method by Lee et al. [LLB23] also shares similarities with our work. The authors quantize instructions for many L-systems and then train a transformer to generate various trees. This allows them to generate new trees by sampling from the transformer.

Our work builds on the notion of Face Extrusion Quad (FEQ) meshes introduced by Pandey et al. [PSB22]. FEQs are quad meshes with only non-intersecting face loops, i.e., no face is contained twice in the same face loop. [PSB22] show that it is possible to decompose FEQ meshes and rebuild the same face loop structures from such recorded sequences. In this paper, we introduce a textual representation for extrusions and use an LLM to learn full extrusion sequences in terms of this representation.

## 3. Face Extrusion Quad (FEQ) meshes

Inspired by Face Extrusion Quad (FEQ) meshes and the associated building methodology developed by [PSB22], we propose a framework for learning sequences of extrusions. The crucial observation is that extrusions can be described in terms of text, which, in turn, can be learnt by an LLM.

In the following, we briefly define the main notions behind FEQ meshes and then describe our method, outlined in Figure 2. Given a quadrilateral mesh whose set of faces is denoted $F$, a *face loop* is a subset, $F_l \subset F$, which forms a contiguous ribbon (cf. Figure 3) such that each face is connected to its neighbors in the loop by opposite edges called *sleeper edges*. The edges along the boundary of the face loop are called *rail edges*. We can traverse a face loop going from face to face across sleeper edges until we come back to the face from which we started. Any pure quad mesh, including non-FEQ meshes, is composed of *face loops*. To be precise, each face of any quad mesh belongs to either one or two face loops. If a face belongs to only one face loop, it is because it is contained twice in that loop which we then consider to be *self-intersecting* as shown in Figure 3 on the right. Non-self-intersecting face loops can be divided into self-adjacent and non-self-adjacent face loops. If two faces that both belong to the loop share an edge which is not a sleeper edge, it is said to be self-adjacent, an example of which is seen in figure 4, where the light blue edges are sleeper edges, and the red edge is a rail edge, which is shared by two faces.

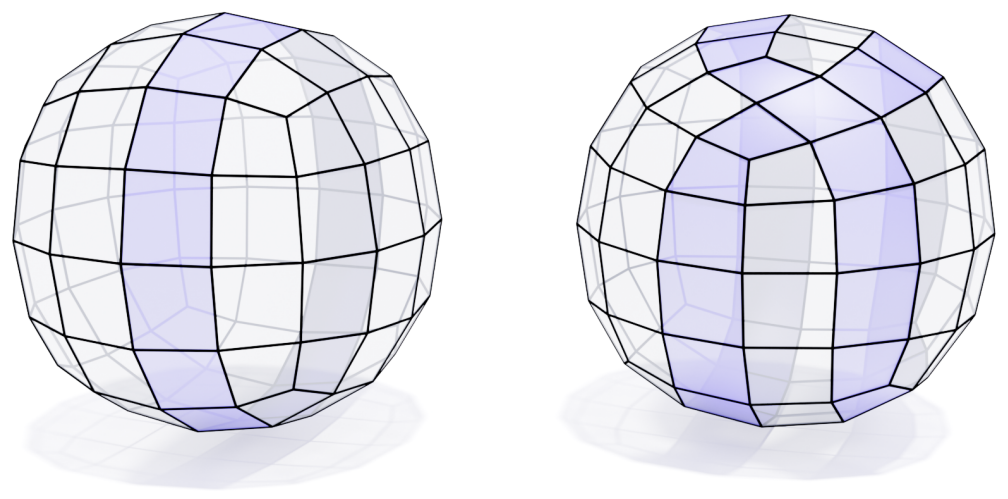

**Figure 3:** *Two meshes, each with a single face loop highlighted in blue. The face loop on the right is self-intersecting.*

An extrusion is an operation on a set of faces, $F_p$, that has disk topology and which we denote the *base patch*. Disconnecting $F_p$ from the rest of $F$ results in two identical boundary edge cycles. The extrusion is completed by adding a face loop, $F_l$, that joins these two boundary cycles, which become its rails, reconnecting $F_p$ to the rest of the mesh. The face set $F_p \cup F_l$ will be denoted the *extended base patch*. Importantly, extrusions are invertible operations, and the inverse is called a *face loop collapse*. The collapse of a face loop removes $F_l$ and stitches the two rail curves to reconnect $F_p$ with the rest of the mesh.

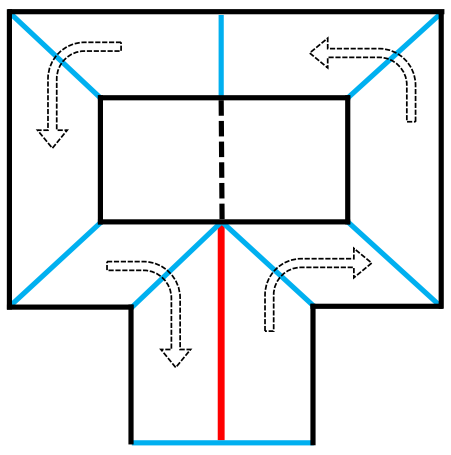

**Figure 4:** *Adopted from [PSB22]: A self adjacent face loop with sleeper edges in light blue, and with direction of traversal through the face loop indicated with arrows.*

A face loop created by a face extrusion is neither self-intersecting nor self-adjacent. This motivates the definition of an *FEQ mesh* as a closed, genus 0 quadrilateral mesh $F$ without self-intersecting nor self-adjacent face loops. In the following, when we refer to face loops, it is understood that these are free of self-intersections and self-adjacency.

### 3.1. From Features to Extrusion Sequences

As [PSB22] point out, we can build a feature of an FEQ mesh by applying a series of extrusions starting from a base patch on the original mesh. Since every extrusion builds on the previous extrusions, there is a clear hierarchical relationship between the extrusions. This relationship can be modeled with a directed acyclic graph (DAG), called the *extrusion graph*, whose nodes represent extrusions, as shown in figure 2. An edge connects two nodes if the base patch of one (the child node) intersects the extended base patch of the other (the parent node). In other words, extrusions have a child-parent relationship if the child extrusion extrudes faces produced by the parent extrusion. Since face loops are produced by extrusions, we will use the word face loop when we focus on the result and extrusion when the operation is in focus.

A key idea in both [PSB22] and our work is that the features of FEQ meshes can be analyzed to identify extrusion sequences that could have generated them. We start by identifying *leaf nodes*. A leaf node is a face loop whose base patch contains no other complete face loops, i.e. $\neg\exists F_l' \subseteq F_l$. An example of such is the tip of a finger on the hand of the human in figure 2. There are usually several leaf nodes to choose from, and we select the one with the smallest base patch area and collapse it. Once a leaf node has been collapsed, it exposes another face loop, which now becomes a leaf. Thus, we can run this process recursively until we arrive at the base patch for the entire feature. This is indicated by the collapse of the face loop selected by the user before the process of the face loop collapses begin. Note that at every collapse, we keep track of which face loops (or nodes) depend on the faces removed by the collapse. This is how we build the DAG that represents the FEQ feature.

Unlike [PSB22], we need to linearize the DAG since it is converted to a text sequence. Topologically sorting the DAG yields a sequence of extrusions that one can follow to rebuild the part of the FEQ that was just collapsed. However, whenever the decomposed feature has a branching structure, there is no unique topological ordering of the extrusions. To make our method robust to the different ways a mesh can be generated, we augment our dataset with random topological orders for each collapsed part of the FEQ. However, only the branching nodes are topologically sorted, and we thus do not permute chains of extrusions, which only have *one* immediate successor and predecessor.

### 3.2. Applying Extrusion Sequences

With a topological ordering of a DAG from an FEQ mesh, we now have a sequence of extrusions to follow that guides the assembling process of the FEQ part we collapsed. We select a set of quadrilateral faces from the mesh $F$ as a base patch and choose a vertex on the boundary, which will determine the initial orientation of the FEQ part. With this, we can start applying the first extrusion in the sequence to extrude the base patch. If the next extrusion in the sequence is a direct continuation of the first extrusion, we keep the orientation from the previous extrusion and use the extruded face set from the first extrusion as the new base patch for the second extrusion. However, if we hit a branch node in the DAG during the assembly process and extrusion $n+1$ in the sequence only depends on a subset of faces in the extended base patch of extrusion $n$, or if extrusion $n+1$ depends on faces from multiple previous extrusions $k, l, m \ldots, n$ $(k, l, m, \ldots, n < n+1)$, we need to select the correct faces from these extrusion(s), which will form the base patch for extrusion $n$. This selection is arguably the most challenging prob-lem we had to solve. If we were to apply extrusion operations only to base patches of exactly the same connectivity as the original collapsed face loops, we would simply need some method for labeling faces. However, as [PSB22] did, we wanted the ability to apply extrusions to arbitrary base patches of varying connectivity. Consequently, after the collapse of extrusion $n+1$, we parametrize the extended base patches of parent extrusion operations $k, l, m \ldots, n$. Next, for each parent extrusion, we find the subset of faces in the extended base patch, which contribute to the base patch of extrusion $n+1$, and store a closed curve $c$ in the parameter domain, which bounds this face subset. Each node in the DAG thus stores the extrusion operation itself as well as a set of closed boundary curves $\{c_k, c_l, c_m, \ldots, c_n\}$, where each curve describes what region of the parameterized parent extended base patch to use.

## 4. Method

Our contribution hinges on the ability to learn extrusion sequences. While it would be possible to build a transformer model from scratch in order to learn the linearized representation of the extrusion DAG, we opted instead for an approach that leverages the ability of modern LLMs to produce structured text.

To perform an extrusion, we need 1) a geometric description of the extruded base patch, 2) references to the previous extrusions we build upon, and, 3) for each of these, a curve that describes the extruded region. This information is converted to textual information for each extrusion in a sequence. We call this representation Text Encoded Extrusions (TEE).

Now, given a training set of different FEQ meshes, we can produce TEE strings for each feature that we decompose from a given mesh. This is sufficient for being able to store and apply extrusion sequences like in [PSB22], but it would not generalize, as each extrusion operation and its associated boundary curve(s) will be unique. In other words, we might hope that a language model could recall an extrusion sequence but not that it would be able to generate something new.

To achieve the ability to generate, we group the extrusions operations together in terms of similarity and discretize the boundary curves. Specifically, each extrusion operation is applied to the same generic base patch, as illustrated in figure 9, and then K-means clustering [PVG*11] on the vertex positions of the meshes is used to cluster similar looking extrusions. For each cluster, the extrusion closest to the mean of the cluster becomes the cluster representative. Afterwards, all the extrusions in the building manuals are substituted with their cluster representative.

### 4.1. Encoding extrusions

In the following, we discuss the specifics of our approach to making extrusion sequences learnable. In [PSB22], the authors delimit the region to be extruded by a closed curve $c$. To allow a curve to cover the same region of a target base patch independently of the geometry and connectivity of the patch, [PSB22] computed a 2D harmonic map of the patch and described the boundary of the faces to be extruded (from the source patch) in this parameter domain. To determine the orientation, a vertex on the boundary was selected as a reference vertex and given the coordinates $(1,0)$, after which the rest of the boundary vertices were mapped to the unit circle.

With the boundary fixed, the interior vertices could be mapped to the plane using the harmonic map, but before computing the map, every quadrilateral face was split into four triangles by inserting a vertex at its center. To apply the extrusion, this parametrization method was also applied to the target patch, and faces whose centroids were inside the closed curve were selected for extrusion.

We adopt this approach and use libigl [JP*18] to compute the harmonic map of the extended base patch. Similar to [PSB22] we only record the curves describing the base patch $F_p$ of extrusion operation $n+1$, if $F_p$ consists of faces that come from multiple extended base patches of extrusions operations $j \in \{k, l, m, \ldots, n\}$, $(k, l, m, \ldots, n < n+1)$, or if $F_p$ is a subset of faces from an extended base patch of only one extrusion operation $j$. In [PSB22], the curves $\{c_k, c_l, c_m, \ldots, c_n\}$, which describe the face set $F_p$ are stored together with the extrusion operation. However, as we need to cluster the extrusion operations, the curves should be disentangled from them. Furthermore, as the LLM predicts a probability distribution over a discrete output space, the curves need to be discretized. Each curve $c$ is a piecewise linear curve, as they bound sets of faces with disk topology. The coordinates, however, are continuous, because it is defined in the domain of the parametrized extended base patches. We therefore introduce a generic extrusion (cf. Figure 5) to discretize the curves. Essentially, the generic extrusion is a high-resolution mesh that covers the unit circle and therefore mimics the harmonic map of a base patch. With this, every point $\mathbf{p}_j \in \mathbb{R}^2$ in a curve $c = \{\mathbf{p}_1, \mathbf{p}_2, \ldots, \mathbf{p}_k\}$ can be mapped to the nearest vertex on the generic extrusion, and the discretized curve $\bar{c}$ thus becomes a sequence of vertex IDs of the generic extrusion $\bar{c} = \{v_k, v_j, \ldots, v_m\}$. This is inherently discrete and, at the same time, provides generalization, as the vertex IDs can occur multiple times in TEE sequences. It also allows us to map back from a vertex ID on the generic extrusion to a 2D point in the harmonic map.

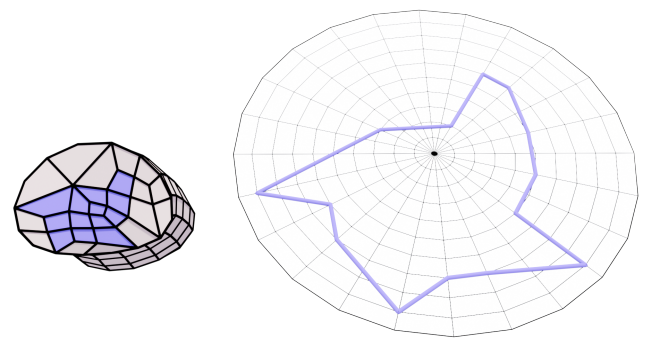

**Figure 5:** ***Generic extrusion**: A set of faces highlighted in blue on an extended base patch to the left, and the boundary curve of the face set mapped to 2D and illustrated on the generic extrusion to the right.*

### 4.2. Fetching the contributing base patches

When we generate a shape by applying each extrusion in the sequence, we also need the TEE string to specify the necessary previous extrusions that will make up the base patch of new extrusion, as intermediate steps will need to be revisited. This happens, if there is a branching node in the DAG obtained from the FEQ mesh. This representation should be discrete to align with the LLM's training scheme.

Specifying the previous extrusions, which we need later in the generation process, is, however, easy: we simply assign each ap-plied extrusion a discrete sequential ID and store the extrusion with its ID in a database.

### 4.3. The Extrusion operation

Similar to [PSB22], we want to be able to store an extrusion operation and apply it to an arbitrary set of faces with disk topology. We therefore adopt their approach and store the extrusion operation during the collapse of a face loop in the following way: Prior to the collapse, the quadrilateral (quad for short) faces of the base patch are triangulated similar to the encoding process of extrusions. Then the 3D-positions $\mathbf{p}_v$ of all the vertices in the triangulated base-patch are stored, after which the face loop is collapsed. The base patch of the collapsed face loop is then flattened by smoothing the interior vertices using Laplacian smoothing, after which a 2D parameterization of the base patch is computed using a Harmonic map [JP*18]. For each vertex in the smoothed triangulated base patch we find it's current position $\tilde{\mathbf{p}}_v$ and then store the displacement vector to its previous position prior to the collapse $\vec{\mathbf{x}_v} = \mathbf{p}_v - \tilde{\mathbf{p}}_v$. The extrusion operation thus contains the source base patch of the collapsed face loop. This source base patch consist of the triangle faces $\{T_1, T_2, \ldots, T_n\}$, and each triangle face $T_i$ consists of vertices $v_k$ with 2D parametric coordinates $(\mathbf{u}_k, \mathbf{v}_k)$ and displacement vectors $\vec{\mathbf{x}}_k$. As an additional attribute, we also store the circumference of the source base patch $L_s$.

To apply the extrusion operation to an arbitrary set of faces with disk topology, which we will denote the target base patch, we follow the procedure above in reverse. First, we smooth the interior vertices of the target base patch using Laplacian smoothing. Then we compute a 2D parameterization of the smoothed target base patch using a Harmonic Map [JP*18], such that we can compare the target base patch with the source base patch through their parameterizations. For each vertex $\tilde{v}_k$ in the parameterized target base patch, we find its enclosing triangle face $T_i$ in the parameterized source base patch using barycentric coordinates. We then interpolate the displacement vectors of the vertices of $T_i$ in the source base patch to find the displacement vector $\vec{\mathbf{x}}$ of $\tilde{v}_k$. With this, we can compute the new position of $\tilde{v}_k$ in the target base patch. However, this will only provide a single estimate of $\tilde{v}_k$'s new position. Consequently, to increase robustness, the position of $\tilde{v}_k$ in the new base patch is computed by mapping the $\vec{\mathbf{x}}$ to the reference frame of each of the boundary vertices in the target base patch and then averaging over all of the estimates of $\tilde{v}_k$'s new position. To account for scale differences between the source and the target base patch, we find the circumference of the target base patch $L_t$, and we then scale the displacement vectors with the ratio $L_t/L_s$.

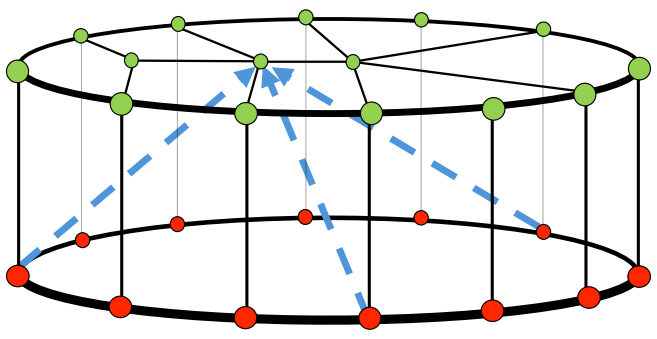

**Figure 6:** ***Extruding the base patch:***. *The extrusion of the vertices in green of the the target base patch is determined by the extrusion vectors (dotted blue vectors) mapped to a local basis at each of the boundary vertices of the base patch (red).*

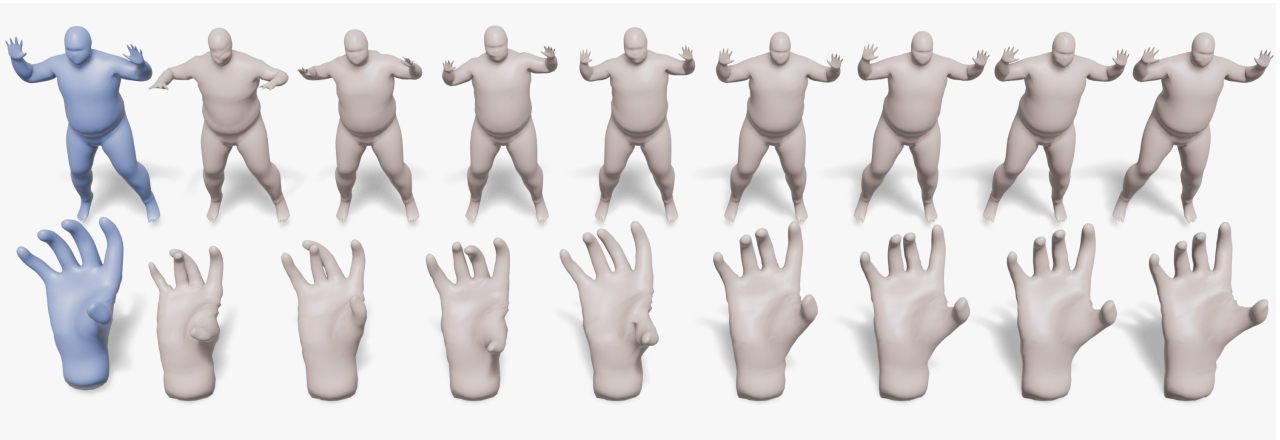

**Figure 7:** ***TEE Reconstruction***: *Clustering of the extrusions in the shape. Ground truth shapes (shown left in blue) and the reconstructions to the right. From left to right, the number of extrusion clusters is:* 1*k*, 5*k*, 10*k*, 20*k*, 30*k*, 40*k*, 50*k* 60*k*. *All shapes reconstructed using quad only buidling methodology.*

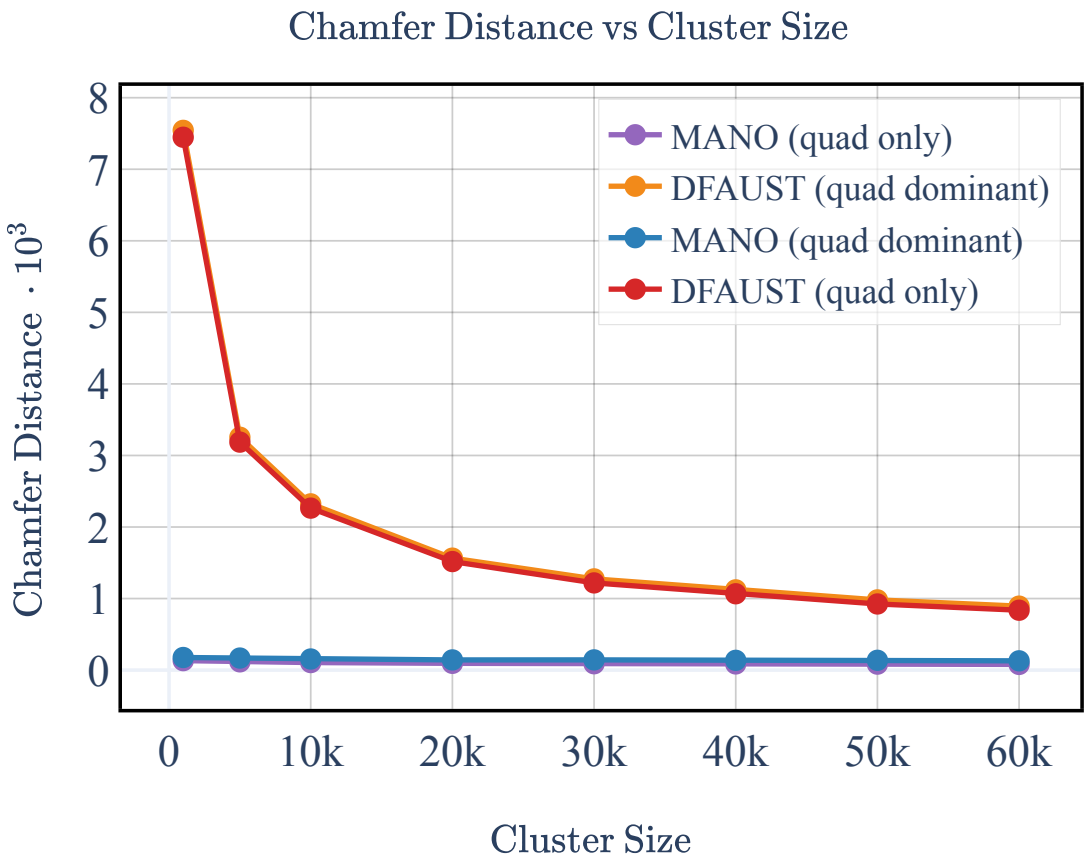

**Figure 8:** ***Reconstruction quality***: *The graph shows the reconstruction quality of the meshes measured as Chamfer Distance as a function of the cluster size. The reconstruction quality is improved when the number of clusters increases, thus allowing for more specific extrusion operations. The performance of both building methodologies is nearly identical.*

Having found the new positions of the vertices $\tilde{v}_k$ in the target base patch, we can detach the patch from the mesh and introduce a new face-loop, which reconnects the elevated target base patch with the mesh.

In Figure 6, an illustration of the application of the extrusion operation is seen. It is clear that different extrusions can yield very different results, as shown in Figure 9. Despite the differences in the extrusion operations, many of them have similar shapes - for instance, the sequence of face loops making up a finger on a hand. This makes it possible to cluster the extrusions, which is crucial for the generalizability of our scheme. The number of clusters directly affects reconstruction quality. For $K$ equal to the number of extrusions in our dataset, we can perform lossless reconstructions. However, this comes at the expense of generalization for the LLM. We choose the number of clusters $K$ such that the reconstruction loss is insignificant. A comparison of the original shape and the shapes reconstructed with different numbers of clusters from our TEE is seen in Figure 7, and a graph depicting the Chamfer distance as a function of cluster size is seen in Figure 8.

As a tradeoff between generalization and visual quality/reconstruction quality, we choose $K = 20,000$ for the DFAUST dataset, $K = 5,000$ for the MANO dataset. For the small mixed dataset of FEQ meshes, we experiment with both $K = 1,000$ and $K = 2,000$ due to its modest size.

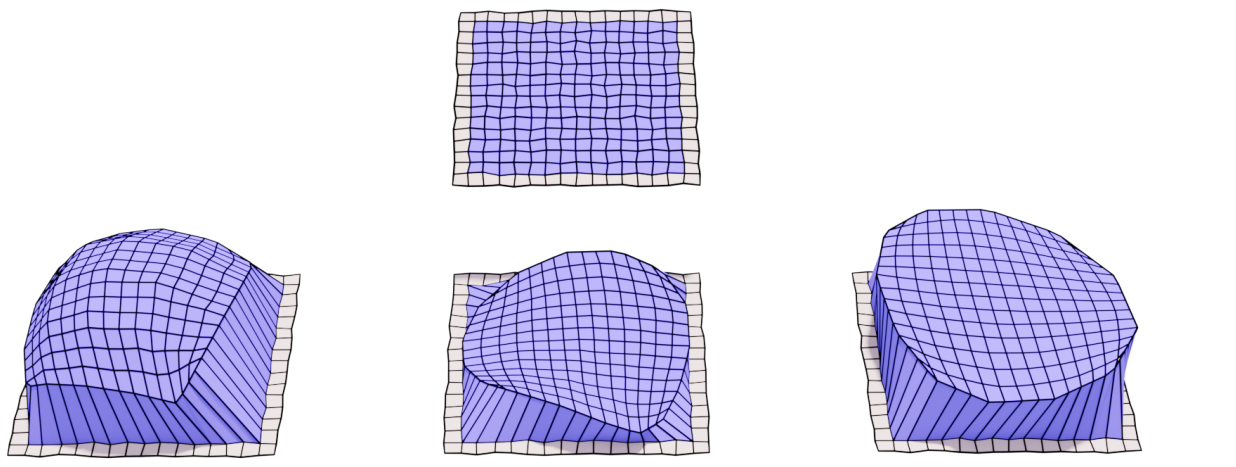

**Figure 9:** ***Applying extrusion operations:*** *Three different extrusion operations are applied to the same generic base patch, creating three different extruded sets of faces.*

### 4.4. Text Encoded Extrusions (TEE)

With the three elements in place: 1) A way to describe to which regions of a base patch that we want to extrude, 2) How to reference the previous extrusions we want to build upon, and 3) The geometric description of an extrusion operation, we now have all the ingredients to generate a mesh. The only remaining thing is the recipe, which we call Text Encoded Extrusions (TEE). This recipe guides our framework in applying the extrusion sequence onto a set of faces with disk topology. An example of TEE is a sequence for the FEQ human mesh in Figure 2 beginning with the structure:

E7762 E18959 E17026 E3334 E16541 E7402 E7402 E8392
E8392 E6594 P9 Re E583 P10 Re ... (1)

The TEE sequence contains a series of commands that are to be followed sequentially. Initially, a user selects a set of connected faces on a mesh, which will function as the base patch. Furthermore, an initial orientation of this base patch is chosen. With this, the generation process determined by the TEE sequence can begin. The first 10 terms in the sequence in (1), E*X*, signals the application of an extrusion operation, where *X* indicates the extrusion cluster to be used. Consequently, the extrusion operation from cluster 7762 is first applied to the user-selected base patch, and this extrudes the set of faces and introduces a face loop. Then the extruded base patch is extruded again by applying the extrusion operation 18959, and a new face-loop on top of the previously generated face loop is created. Afterwards, the extrusion operation 17026 is applied and creates a third face loop. This process continues until the 10'th extrusion 6594 is applied. At this point in the sequence, we hit a branching node, and the extended base patch just created by the extrusion operation 6594 will serve as base patch for multiple future extrusion operations. We thus store the current state in the mesh and assign it the ID *P9*, because it is the ninth extrusion operation in the TEE sequence (using 0-indexing). We then continue to *Re*, which does not have any geometric implication, but only acts as a special token to emphasize that this state will be needed later on. The next command is then another extrusion operation 583, and this is applied on top of last recorded base-patch before the command *P9*, which was the base-patch extruded by operation 6594. We then encounter a new branching node in the sequence, and assign it the ID *10*. We then also remember this state *Re* and move on with the sequence until there are no more commands to be followed.

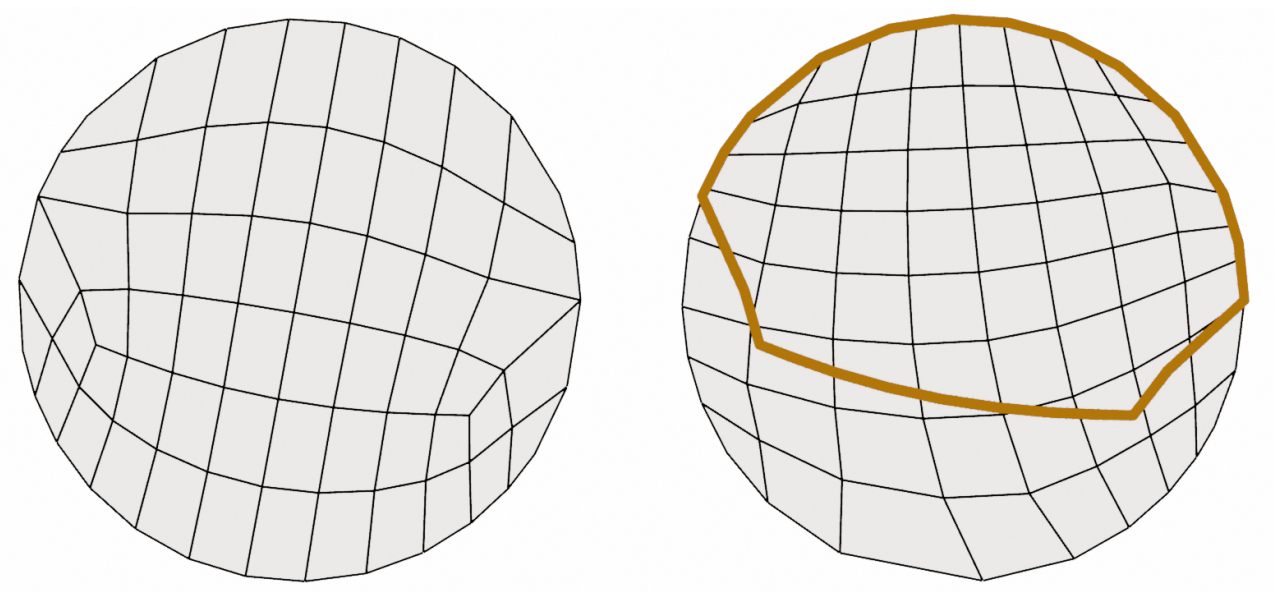

**Figure 10:** ***Face selection set:*** *Given a boundary curve from the parameterized source base patch on the left, it is a nontrivial task to select the correct set of faces on the parameterized target base patch on the right, as some may only be partially included within the closed curve*

## 5. Implementation

Finding the appropriate faces on a mesh to apply an extrusion operation can be challenging, if the connectivity of the base patch in the beginning of a sequence of extrusions is very different from the connectivity of the source base patch of the sequence. The difference in connectivity causes the 2D parameterization of the target base patch to differ too much from the 2D parameterization of the source base patch on which the extrusion operation was obtained, thus making it a non-trivial problem to find the appropriate faces, as they may only be partially within the closed loop as illustrated in Figure 10. Consequently, we have developed two different building methodologies: A pure quadrilateral one and a quad-dominant approach. What sets the latter building methodology apart is the ability to cut through the base patch according to the curve denoting the faces needed for the extrusion. If a point on the curve falls within a quadrilateral face, we insert a vertex in the face, and if a curve segment intersects an edge in the patch, we split the edge and insert a new vertex at the intersection point. These changes might result in polygonal faces with degree more than 4. We therefore triangulate these faces using a Delaunay triangulation using Triangle [She96] while maintaining the edges making up the loop. No matter the building methodology, the TEE sequence is the same. The result however can be quite different as illustrated in figure 11, where the shape on the left was reconstructed using the pure quadrilateral building methodology, and the three meshes in the middle and to the right with the quad dominant building methodology. While it is preferable to use the pure quadrilateral building methodology, as it aligns with the way that designers create meshes, the quad-dominant approach provides robustness and ensures that the TEE sequence can be applied to a base patch of arbitrary connectivity and geometry and still generate the desired shape.

For the pure quadrilateral method, we use a method inspired by the notion of flux to select a set of faces with disk topology, whose boundary curve matches the closed loop. First, we sample $N$ equally spaced points $p_i \in \mathbb{R}^2$ on each of the line segments of the curve, and for each of these points we associate a vector $\mathbf{v}$, which is the normal of the line segment. Then we sample another $N$ equally spaced points $q_k \in \mathbb{R}^2$ on each half-edge $e$ in the parameterized base patch. With this, we can compute a weight $w(e)$ for each half-edge $e$ given by:

$$w(e) = \sum_{k=1}^{N} -\frac{\mathbf{n}_e}{\|\mathbf{n}_e\|} \cdot \frac{\mathbf{v}(q_k)}{\|\mathbf{v}(q_k)\|} + \alpha \cdot d(q_k)^2 \quad (2)$$

Here $\mathbf{n}_e$ is the normal of the edge $e$, $\mathbf{v}(q_k) = v_{i*(q)}$ is the normal of the line segment that is closest to $q_k$ given by the index $i^*(q_k) = \arg\min_i \|q_k - p_i\|$, and $d(q_k) = \|q_k - p_{i^*(q_k)}\|$. With this weight function, we want to assign numerically large values to half-edges that are parallel to the curve and numerically low values to half-edges that are orthogonal to the curve. Once all half-edges have been assigned a weight, we sort them according to the lowest weights, and create a Minimum Spanning Tree (MST) with them, hence why we use the negative dot product between the normals of the edge and the closest line segment. To distinguish between parallel edges close to the curve and further away from the curve, we add a penalty term $\alpha \cdot d(q_k)^2$ and use $\alpha = 50.0$, which was determined experimentally. After having created the MST, we go through each of the edges that are not in the MST and detect, which cycle they create. We compare each cycle with the curve by computing the Dynamic Time Warping (DTW)-distance [Sen08] and then select the cycle with the smallest DTW-distance. This cycle will be the sequence of edges, which resembles the curve the most. It will then automatically enclose a set of faces, and these will be part of the base patch.

## 5.1. LLM

To model our TEE, we use the Llama 3.2 1B model [DJP*24]. This model was chosen for its relatively large context window and its small number of parameters, which allow us to fine-tune it on our dataset. Since TEE is a text-based representation, fine-tuning an LLM can be done without modification of the LLM. We did however decide to add a extra tokens to the vocabulary, as TEE does not resemble ordinary natural language. These extra tokens were based on the frequency of appearance in the text dataset. This was however only a slight modification, as the core of our framework is not the neural network, but instead the geometric extrusion language, which generates a shape. Our method is thus agnostic to the LLM that is used.

## 6. Dataset

A dataset of FEQ meshes is essential for our method. These meshes must be 2-manifold, genus-0, quadrilateral, and free of self-intersecting and self-adjacent face loops. Such meshes are difficult to obtain in practice. Consequently, we use a combination of datasets, including surface meshes from the Hexalab dataset [BTP*19], a database of papers on hexahedral meshes. We filter out shapes that have triangles, self-intersecting loops, or a genus greater than 0. The authors of [PSB22] also provided us with the FEQ meshes from their work. These combined however still only make a modest dataset. Consequently, to diversify our FEQ meshes, we also created two additional FEQ datasets based on the MANO dataset (the right hands in the dataset) [RTB17] and the DFAUST dataset (the upper parts of the bodies) [BRPMB17]. We use an FEQ mesh of a hand and map the connectivity of this mesh to the triangle meshes in the MANO dataset using a bijective map computed between the FEQ hand mesh and one triangle mesh from the MANO dataset [SPK23]. Similarly, we modify a human mesh from [CG6] to obtain an FEQ mesh. We then map this modified mesh to the humanoids from the DFAUST dataset using, again, a bijective mapping computed between these two meshes [SPK23]. Since the DFAUST set contains time series of humans performing activities, we use a subset of meshes that are temporally distant to avoid including shapes that are very similar.

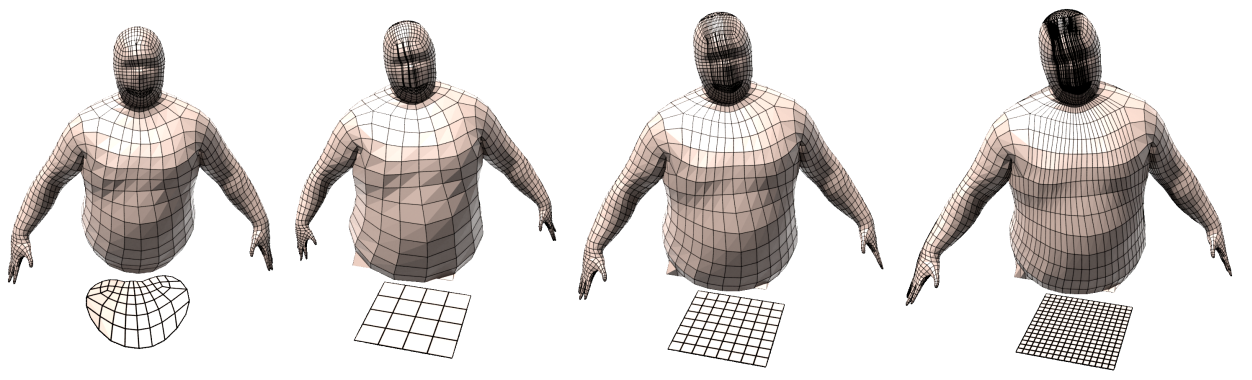

**Figure 11:** ***Versatility of TEE**: On the left, the torso is reconstructed from the standard body base patch, whereas high-res square patches of increasing resolution are used on the right. From left to right, the ratio between quadrilateral faces and triangle faces of meshes are:* 2444/0, 2132/342, 3300/660, 5400/1684.

In total, we operate on three datasets: The right hands from the MANO dataset, a subset of the DFAUST dataset, and a small highly diverse database of FEQ meshes generated using the skeleton-based conversion in [PSB22] as well as surfaces extracted from hex meshes. We train a Llama 3.2 model on each of these datasets.

## 7. Results

In the following we first show that TEE is a versatile and flexible shape modeling language. Afterwards, we demonstrate the shape synthesis capability of the fine-tuned LLM on generating novel TEE sequences, and its ability to complete features on shapes not in the training dataset. Lastly, we show how the MeshXL method compares to our approach in terms of synthesizing shapes.

## 7.1. TEE versatility

To demonstrate the reconstruction capability of TEE and its invariance to the base patch, we apply the same TEE sequence of a DFAUST human onto four different base patches, as seen in Figure 11. For the original base patch shown to the left in the figure, we are able to construct the mesh using the quadrilateral only method. If the base patch differ too much from the original one, as is the case with the square base patches of various resolution, we can simply switch to the quad dominant building methodology, and still generate the mesh in the way that we intended it. Despite small deviations, it is clear that all meshes are geometrically very similar.

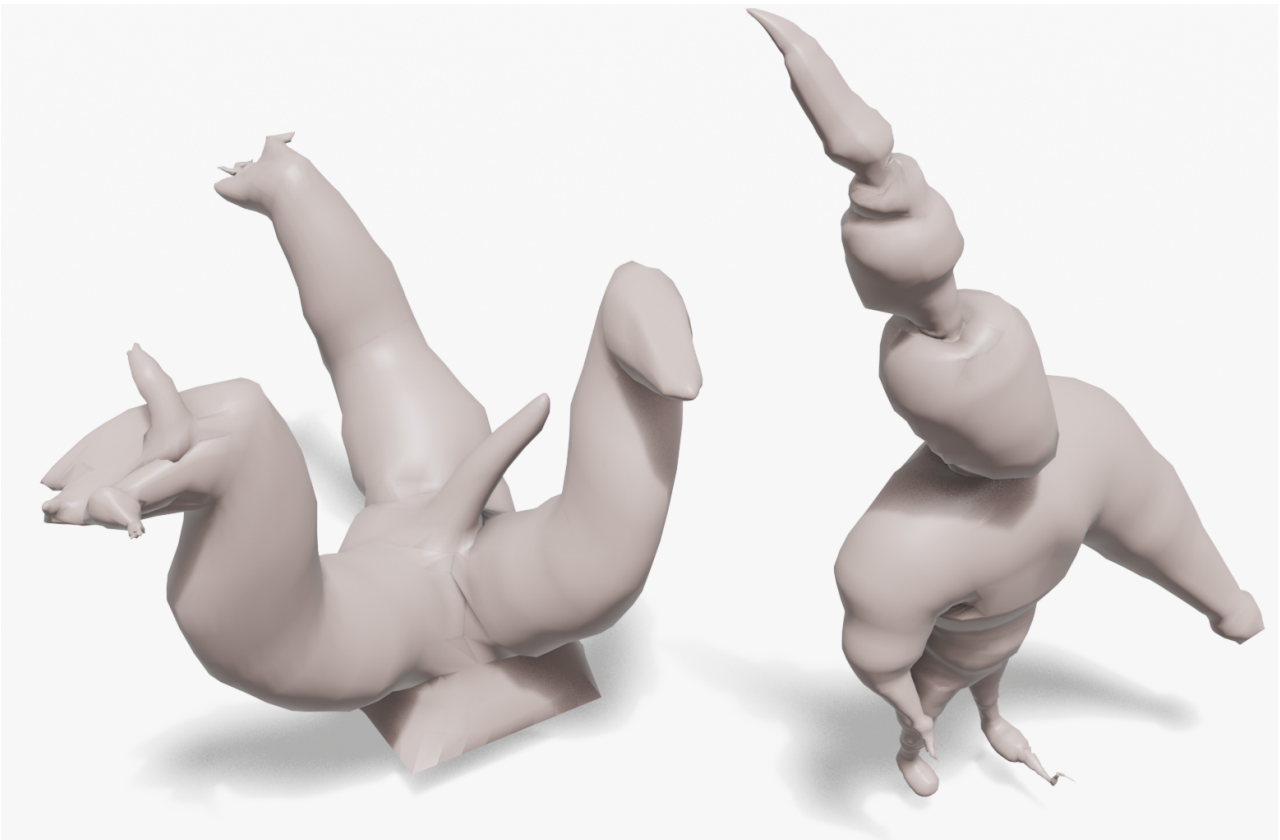

**Figure 12:** *Hallucinations of shapes not present in our diverse dataset of FEQ meshes. The ratio between quadrilateral faces and triangle faces of the meshes is:* $7933/176$ *and* $19686/492$ *for the left and right respectively.*

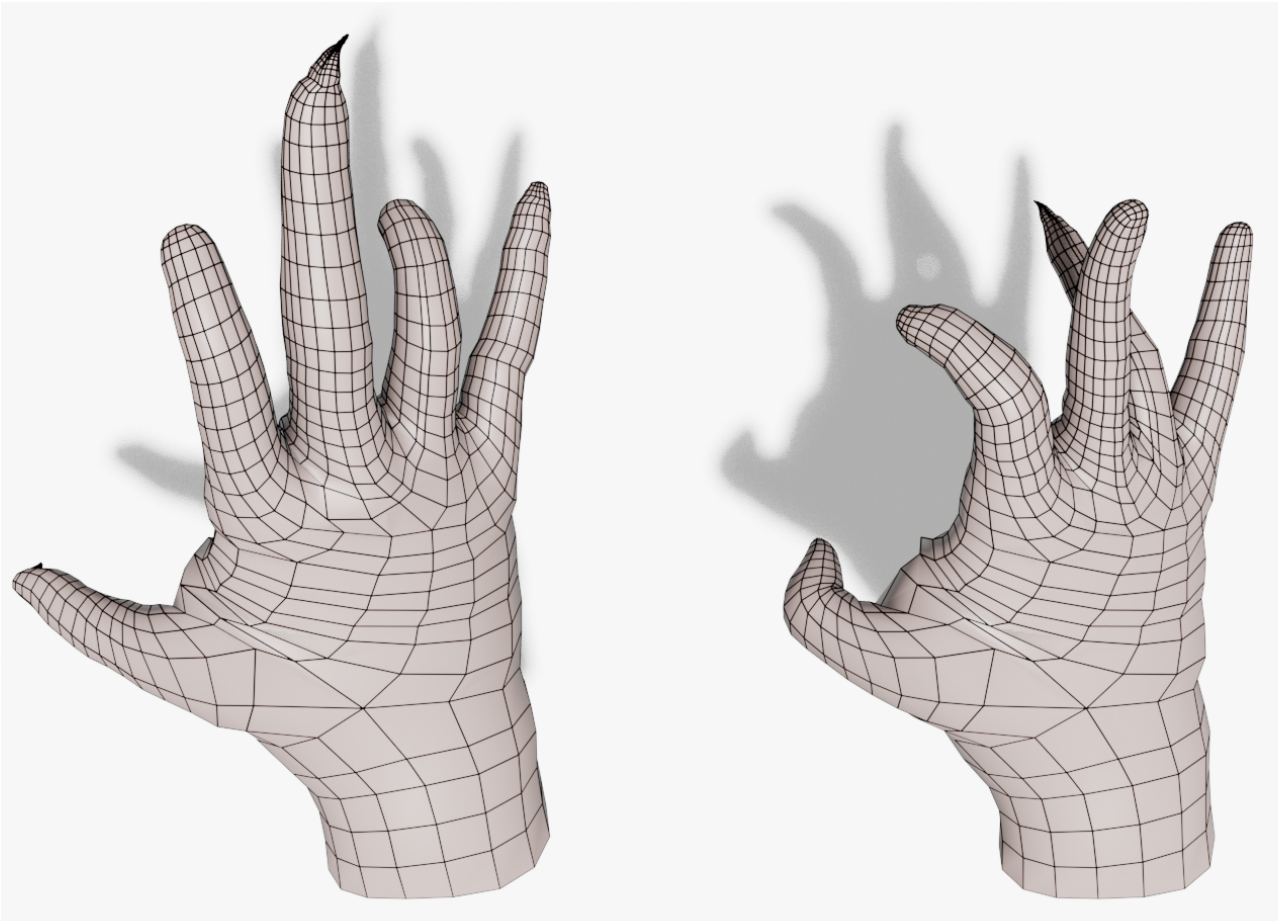

**Figure 13:** *TEE sequences of hands with additional features. The ratio between quadrilateral faces and triangle faces of the meshes is:* $2173/0$ *and* $3467/0$ *for the left and right respectively.*

If we however wish to let the design of the base patch influence the structure of the generated mesh, we can choose to do by using the quadrilateral only building method. In Figure 15 we demonstrate this, by decomposing the hat of an elf mesh and applying the resulting TEE sequence to various odd-shaped base patches using the quadrilateral only building method. The shape of the base patches then influence the size and the structure of the generated meshes.

With the results in Figure 9, 11 and 15 we thus demonstrate the versatility of TEE and show that the space of possible shapes to generate is spanned by the shape and size of the base patch, the number of extrusions operations in a cluster and the building methodology used.

### 7.2. Shape synthesis

With the fine-tuned LLM we are able to sample new shapes unconditionally, and in Figure 1, we demonstrate this shape synthesis capability. We use the fine-tuned LLM to generate different TEE sequences of DFAUST upper bodies, and we then create new shapes by applying the TEE sequences onto a mix of lower bodies used as base patches. We observe that the LLM is capable of learning meaningful TEE sequences.

In Figure 14, we also show its editing capability, where we have removed the fingers on a DFAUST human and substituted them with TEE sequences generated by the LLM.

In Figure 12, we show two meshes generated by the LLM from our diverse dataset which do not resemble any single model and thereby demonstrate that the method has the ability to produce results that combine features from different parts of the training data.

Our method does however also fail at points, which is seen in Figure 13, where it has generated TEE sequences of MANO hands that were not stopped at the right time–resulting in hands with pointy fingers.

### 7.3. Feature completion

We demonstrate feature completion in Figure 16, as the LLMs capability to condition on part of the mesh and suggest the missing features. We find three DFAUST meshes which are not in the training dataset and remove an arm. We then decompose the incomplete meshes and find the DAG for each of them, which we topologically sort. The linear orderings of the DAGs are then modified such that the missing feature would appear last, if the DAGs were complete. We then convert the orderings to TEE sequences and use those as prompts in the LLM. For each of prompt, we score the LLM output using average log probability, and among the highest scoring/most probable TEE outputs, we select the among the best ones.

### 7.4. Comparison with other methods

As previously stated, our method is orthogonal to prior work on LLM-based mesh generation, as we operate at the component level rather than the primitive level. This allows us to represent shapes with many more faces, which exceeds the limitations of the other works. We have compared our results against MeshXL [CCP*24], which is one of the most recent transformer based methods.

To be able to generate results for MeshXL, we decimated the meshes to 800 triangle faces and fine-tuned it on the DFAUST and MANO dataset. A qualitative comparison of the generated meshes is seen in Figure 18 and in Figure 17. Even when the meshes are decimated, MeshXL still generates meshes with missing polygonal faces.

## 8. Discussion and Limitations

By formulating 3D mesh generation as text synthesis, we are able to bring the ability of large language models to generate structured text to bear on this challenging problem. Moreover, by generating extrusion sequences rather than individual mesh primitives, we have decoupled the output mesh resolution from the complexity of the string from which it is produced. This allows us to generate very large meshes using our methods, something that mesh-based generative methods otherwise tend to struggle with.

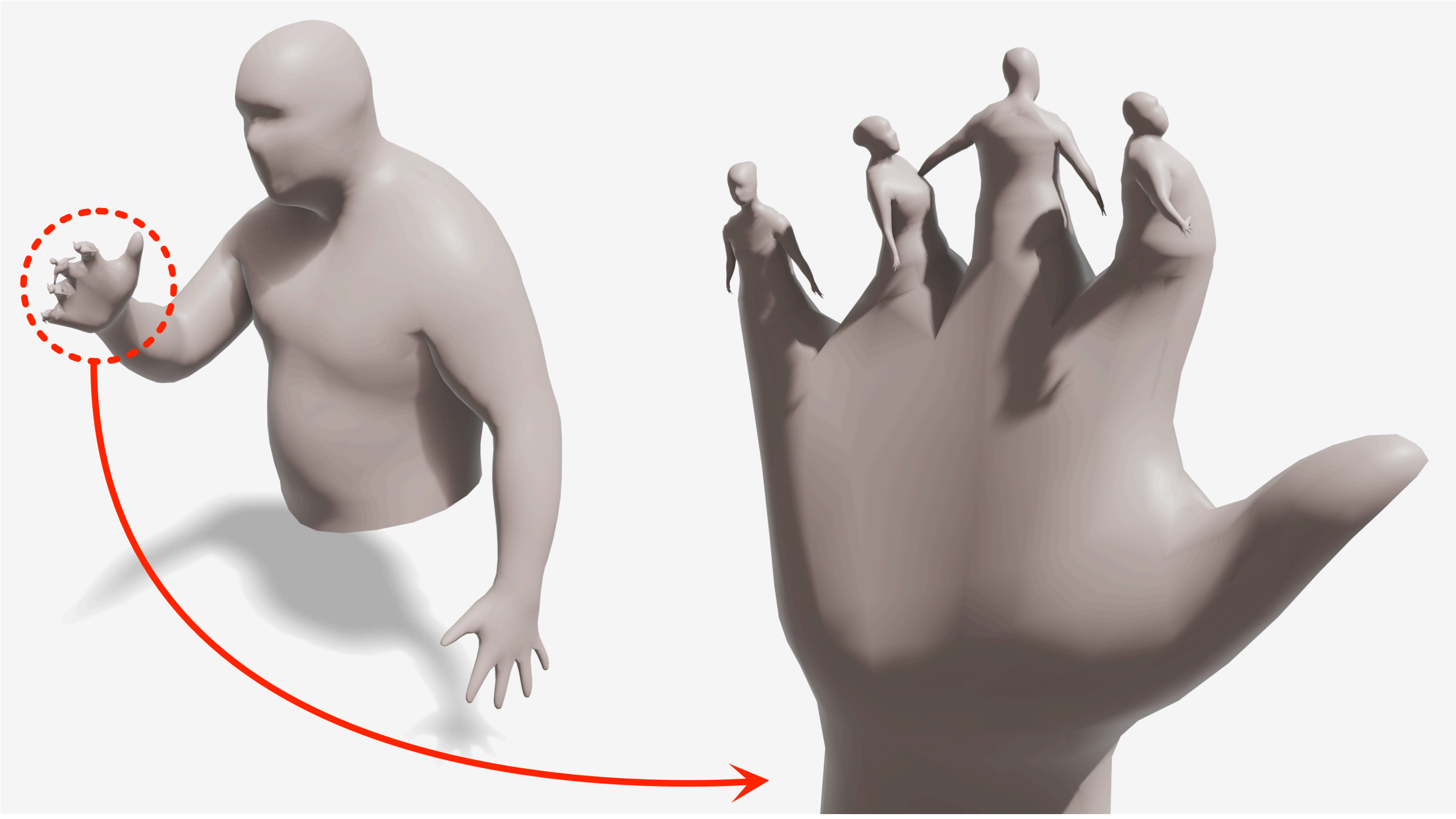

**Figure 14:** *Examples of LLM generated TEE sequences generated on the finger tip of a human mesh. Smoothing was used to deal with the jump in resolution as a post processing step.*

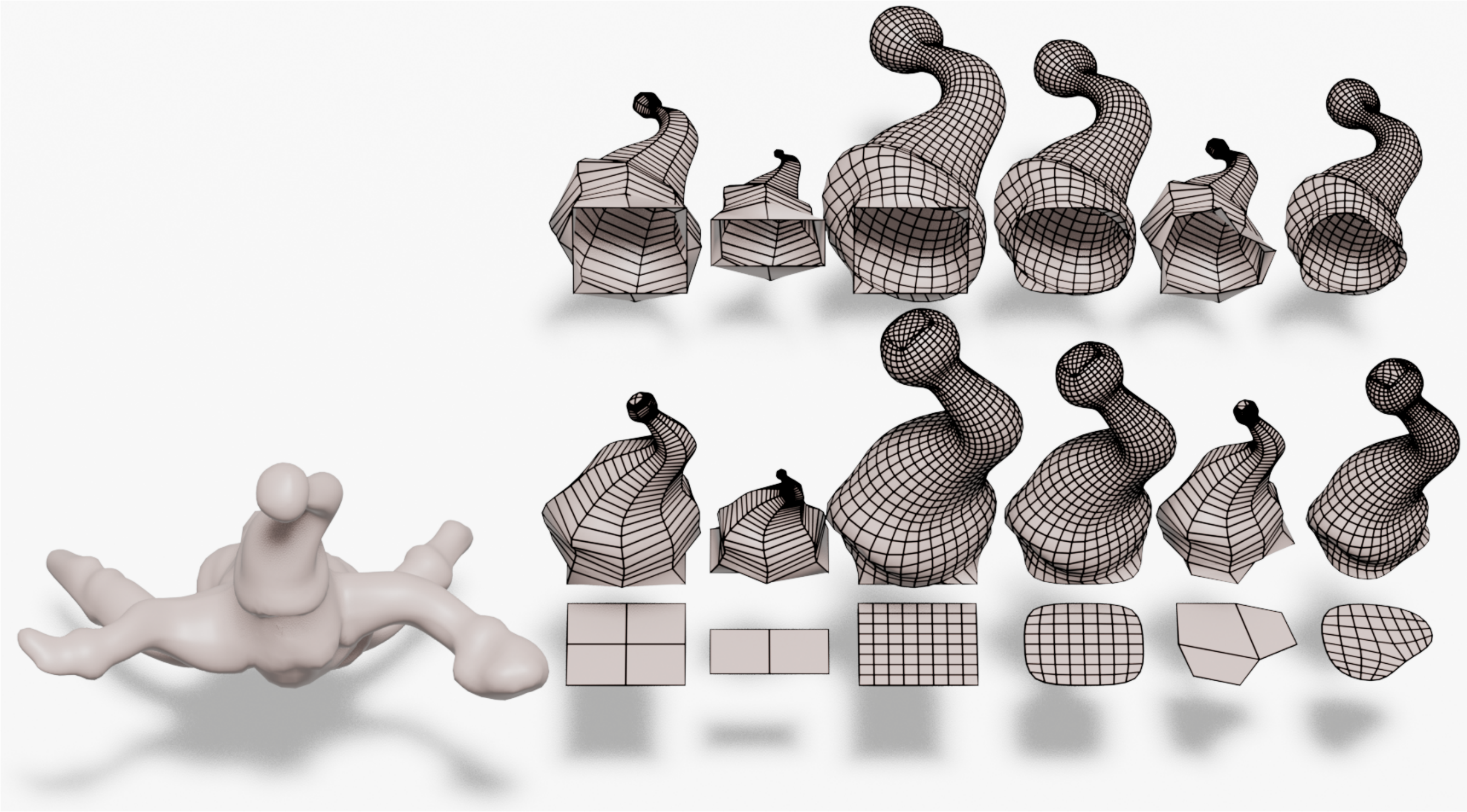

**Figure 15:** *Examples of the same TEE sequence obtained from decomposing the hat of the elf applied to oddly shaped base patches*

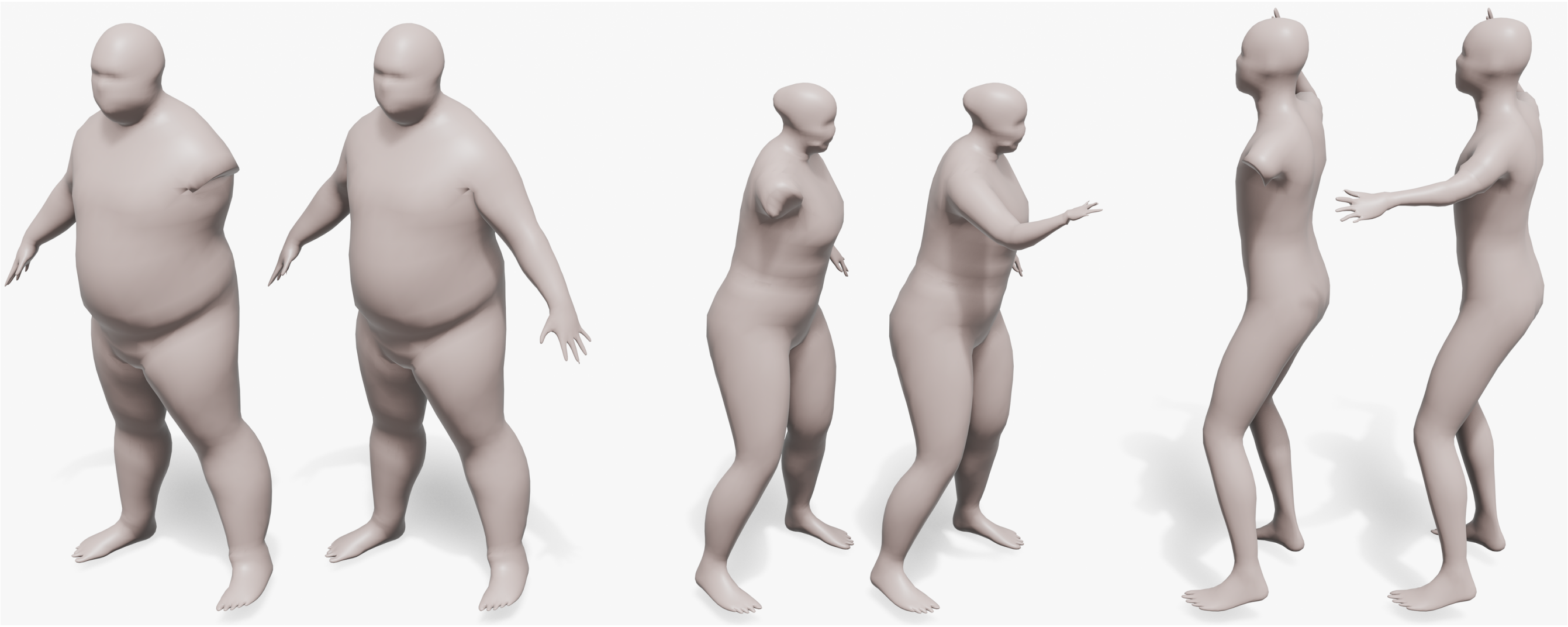

**Figure 16:** *Examples of LLM feature completions, where each pair of meshes consists of a mesh to left, which acted as a 'prompt' to the LLM, and the mesh to the right, with the feature, was the result.*

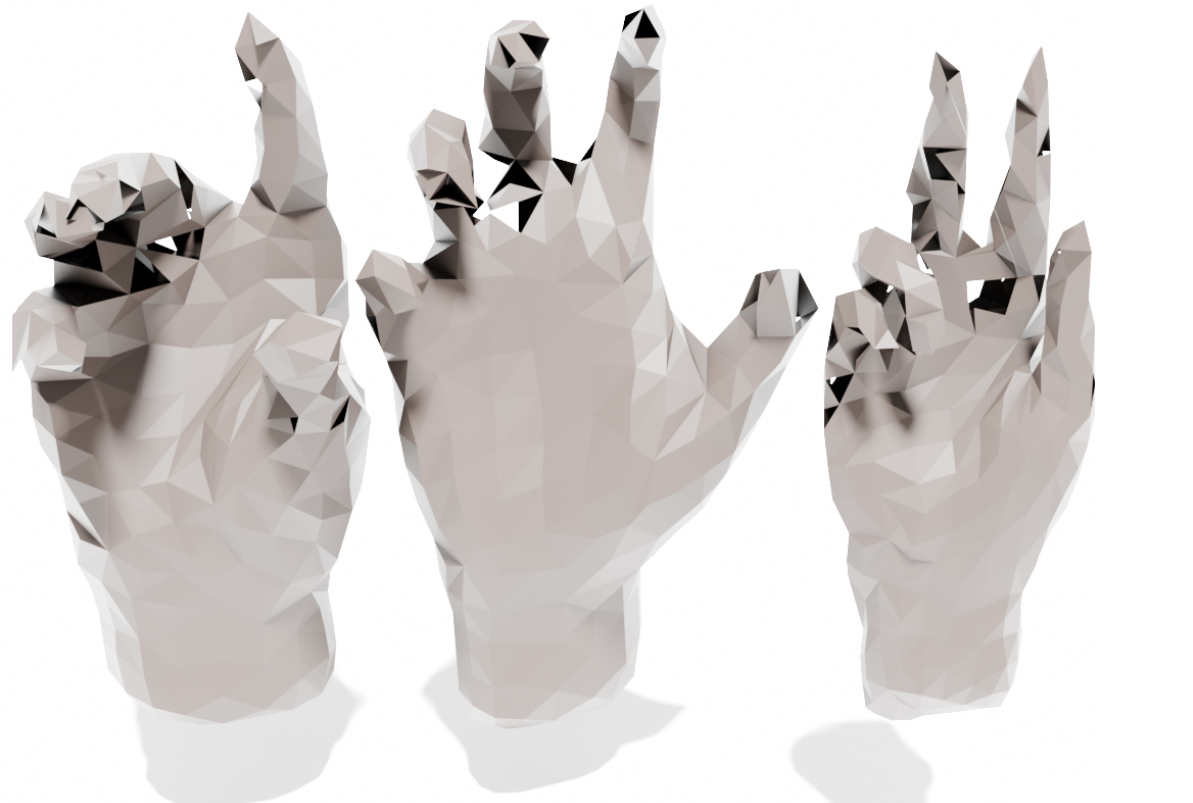

**Figure 17:** ***MeshXL**: Three hands generated using MeshXL fine-tuned on the MANO dataset*

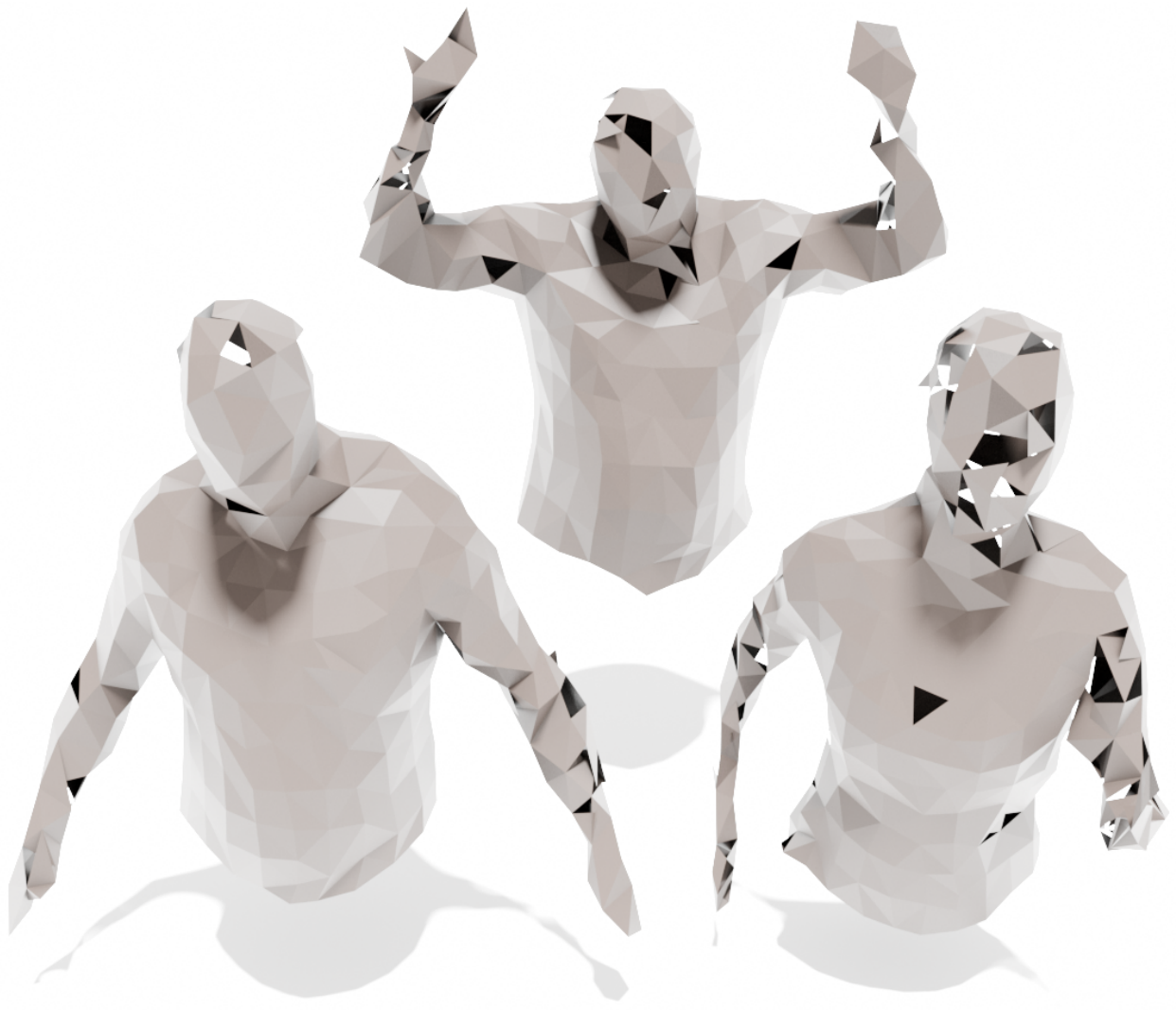

**Figure 18:** ***MeshXL**: Three humans generated using MeshXL fine-tuned on the DFAUST dataset*

We believe this approach is effective, but it does come with challenges and some limitations. First, our method assumes that the input 3D shapes are FEQ-meshes of sphere topology. These are not the most common type of 3D data. Second, many GPU hours are needed in order to effectively leverage the capabilities of LLMs, even with a small fine-tuned model.

However, both of these problems could be overcome in the near future. Regarding the availability of training data, procedural methods could potentially be used to synthesize large amounts of highly varied FEQ models. The current explosive growth in compute resources bodes well for the feasibility of training this type of models.

In its current state, the method is able to synthesize structures based on just a base patch and to generate the missing parts of existing meshes. Going forward, we are hopeful that with a much bigger and more diverse set of training examples, it would be possible to use this type of model to assist artists by allowing them to efficiently add features to 3D models. If labeled data is used for training, it may be possible to guide such generation using images or text prompts.